\documentclass[12pt,a4paper]{article}
\usepackage{jcappub}

\pdfoutput=1 

\usepackage{tikz}

\def\be{{\boldsymbol e}}
\def\bk{{\boldsymbol k}}
\def\bx{{\boldsymbol x}}

\def\cA{{\cal A}}
\def\cL{{\cal L}}
\def\cP{{\cal P}}

\def\ri{{\rm i}}
\def\rr{{\rm r}}
\def\rmm{{\rm m}}
\def\rP{{\rm P}}

\def\max{{\rm max}}

\def\abs#1{{\left| #1 \right|}}

\title{Model-independent constraints in inflationary magnetogenesis}

\author[a,b]{Yuri Shtanov}
\author[c]{and Mykhailo Pavliuk}

\affiliation[a]{Bogolyubov Institute for Theoretical Physics, \\ Metrologichna St.\@ 14-b, Kiev 03143, Ukraine} %
\affiliation[b]{Astronomical Observatory, Taras Shevchenko National University of Kiev, \\ Observatorna St.\@ 3, Kiev 04053, Ukraine} %
\affiliation[c]{Department of Physics, Taras Shevchenko National University of Kiev, \\ Academician Glushkov Ave.\@ 2, Kiev 03022, Ukraine}%

\emailAdd{shtanov@bitp.kiev.ua}
\emailAdd{pavliukconnection@gmail.com}

\abstract{We derive a simple model-independent upper bound on the strength of magnetic fields obtained in inflationary and post-inflationary magnetogenesis taking into account the constraints imposed by the condition of weak coupling, back-reaction and Schwinger effect. This bound turns out to be rather low for cosmologically interesting spatial scales. Somewhat higher upper bound is obtained if one assumes that some unknown mechanism suppresses the Schwinger effect in the early universe. Incidentally, we correct our previous estimates for this case.}


\arxivnumber{2004.00947}

\begin{document} 
\maketitle
\flushbottom


\section{Introduction}

Origin of magnetic fields, observed in different objects and on different spatial scales in the universe \cite{KF}, remains to be unclear (for reviews, see \cite{Grasso:2000wj, Widrow:2002ud, Giovannini:2003yn, Kandus:2010nw, Durrer:2013pga, Subramanian:2015lua}). One of the widely discussed possibilities is that they were generated at the inflationary stage; this naturally explains their large coherence length, which can be comparable to the size of the large-scale structure \cite{Tavecchio:2010mk, Ando:2010rb, Neronov:1900zz, Dolag:2010, Taylor:2011}.  Numerous attempts at realising inflationary magnetogenesis were made by introducing non-trivial couplings of other fields to the electromagnetic field that break the conformal invariance of the latter. A general class of models that we consider here is described by a gauge-invariant action of the form, pioneered in \cite{Turner:1987bw, Ratra:1991bn},
\begin{equation}\label{Lem}
	\cL_{\rm em} = - \frac14 I^2 \left( F_{\mu\nu} F^{\mu\nu} + f \tilde F_{\mu\nu} F^{\mu\nu} \right)  \, ,
\end{equation}
where $\tilde F_{\mu\nu} = \frac12 \epsilon_{\mu\nu}{}^{\alpha\beta} F_{\alpha\beta}$ is the Hodge dual of the electromagnetic strength tensor $F_{\mu\nu}$, and $I$ and $f$ are non-trivial functions of the cosmological time due to their possible dependence on the background fields such as the inflaton or the metric curvature.

Inflationary magnetogenesis theory is confronted by the issues of back-reaction and strong gauge coupling \cite{Demozzi:2009fu, Urban:2011bu, BazrafshanMoghaddam:2017zgx, Sharma:2018kgs}, and by the Schwinger effect \cite{Kobayashi:2014zza, Sharma:2017eps, Stahl:2018idd, Kitamoto:2018htg, Sobol:2018djj, Sobol:2019xls, Gorbar:2019fpj, Sobol:2020frh, Domcke:2018eki, Domcke:2019qmm, Shakeri:2019mnt} that all reduce its efficiency. Various models of evolution of $I$ and $f$ were tried in the literature to overcome these difficulties. In this paper, assuming the condition of weak coupling $I \gtrsim 1$, we would like to provide general model-independent upper limits imposed by the back-reaction and Schwinger effect on the possible strength of magnetic field on a particular comoving spatial scale. Our investigation is similar in spirit to \cite{Demozzi:2009fu, Urban:2011bu} except that we will not impose a priori ansatz for the evolution of the couplings and also take into account the Schwinger effect. 

Given a particular comoving spatial scale, characterised by wavenumber $k$, one can start amplifying electromagnetic field on that scale well before its Hubble-radius crossing during inflation. One possible way of achieving this was demonstrated in our previous publications \cite{Shtanov:2019civ, Shtanov:2019gpx} and will be reviewed below. It suffices to consider evolution of the chiral coupling $f (\eta)$ in (\ref{Lem}) which is linear in conformal time $\eta$ in some time interval. Then, only the modes of one helicity in a narrow band with $k \sim |f' (\eta)|$ will be exponentially amplified.  It was mistakenly claimed in our previous papers \cite{Shtanov:2019civ, Shtanov:2019gpx} that this amplification alone could produce large-scale magnetic fields of sufficiently large strengths today. What was overlooked in \cite{Shtanov:2019civ, Shtanov:2019gpx} is that the peak of electromagnetic field in this process is attained by the moment of Hubble-radius crossing during inflation. The subsequent inflationary dilution of electromagnetic fields then makes them quite small by the end of inflation.  

Hence, it is necessary to amplify the fields also outside the Hubble radius during and after inflation.  At all stages, one should take into account the constraints imposed by back-reaction and Schwinger effect.  This will be the subject of the present paper, in which a simple model-independent upper bound for the final magnetic field is derived. This upper bound appears to be quite low for the prospects of inflationary magnetogenesis.  Somewhat higher upper bound is obtained if one assumes that some unknown mechanism suppresses the Schwinger effect in the early universe.

\section{The setup}

We work in a spatially flat cosmological model with the metric
\begin{equation}
	ds^2=a^2 (\eta) \left( d\eta^2 - \delta_{ij} d x^i d x^j \right) \, ,
\end{equation}
where $\eta$ is the conformal time, and $a$ is the scale factor.  Adopting the longitudinal gauge $A_0 = 0$, $\partial^i A_i = 0$ for the vector potential, from (\ref{Lem}) one obtains the equation satisfied by the transverse field variable $A_i$:
\begin{equation} \label{eqAi}	
	\left( I^2 A_i' \right)'  - I^2 \nabla^2 A_i + \left( I^2 f \right)' \epsilon_{ijk} \partial_j A_k = a^2 j_i \, , 
\end{equation}
where $j_i$ is the electromagnetic current density, $\epsilon_{ijk}$ is the spatial Kronecker alternating tensor with $\epsilon_{123} = 1$, and the prime denotes the derivative with respect to the conformal time $\eta$. Assuming the Ohm's law in a highly conducting medium with conductivity $\sigma$, we have
\begin{equation}\label{Ohm}
	j_i = \sigma E_i = - \frac{\sigma}{a} A_i' \, .
\end{equation}

In the spatial Fourier representation, and in the constant (in space and time) normalized helicity basis $\{ e^h_i (\bk) \}$, $h = \pm 1$, such that $\ri \bk \times \be^h = h k \be^h$, we have $A_i = \sum_h \cA_h e^h_i e^{\ri \bk \bx}$.  Then equations (\ref{eqAi}), (\ref{Ohm}) for the helicity components $\cA_h$, imply
\begin{equation} \label{eqAh}	
	\left( I^2 \cA'_h \right)' + a \sigma \cA_h' + \left[ I^2 k^2 + h k \left( I^2 f \right)' \right] \cA_h = 0 \, ,
\end{equation}
or, equivalently,
\begin{equation}\label{eqAh1}
	\left( I \cA_h \right)'' + \frac{a \sigma}{I^2} \left( I \cA_h \right)' + \left[ k^2 + h k \frac{\left( I^2 f \right)'}{I^2} - \frac{I''}{I} - \frac{ a \sigma I'}{I^3} \right] I \cA_h = 0 \, .
\end{equation}

The spectral densities of quantum fluctuations of magnetic and electric field are characterized, respectively, by the standard relations\footnote{These are electromagnetic fields `felt' by the charged matter, in the Lagrangian of which we assume the standard couplings of the form $\left| \left( \partial - \ri q A \right) \psi \right|^2$, where $\psi$ is the matter field and $q$ is its electric charge. We work in the system of units $\hbar = c = 1$.}
\begin{align} \label{specB}
\cP_B &\equiv \frac{d \left\langle B^2 \right\rangle}{d \ln k} = \frac{k^4}{8 \pi^2 a^4} \sum_h \left| \cA_h (\eta, k) \right|^2 \, ,  \\
\cP_E &\equiv \frac{d \left\langle E^2 \right\rangle}{d \ln k} = \frac{k^4}{8 \pi^2 a^4} \sum_h \left| \frac{\cA'_h (\eta, k)}{k} \right|^2 \, , \label{specE}
\end{align}
in which the amplitude of the vector potential is normalized so that $I \cA_h \sim e^{- \ri k \eta}$ as $\eta \to - \infty$.  The factor in front of the sums in (\ref{specB}) and (\ref{specE}) is the spectral density of the normal vacuum fluctuations in each mode at the physical wavenumber $k/a$.  The (nonrenormalized) spectral energy densities per logarithmic interval of wavenumbers are given by
\begin{equation}
\frac{d \rho_B}{d \ln k} = I^2 \cP_B \, , \qquad \frac{d \rho_E}{d \ln k} = I^2 	\cP_E \, .
\end{equation}

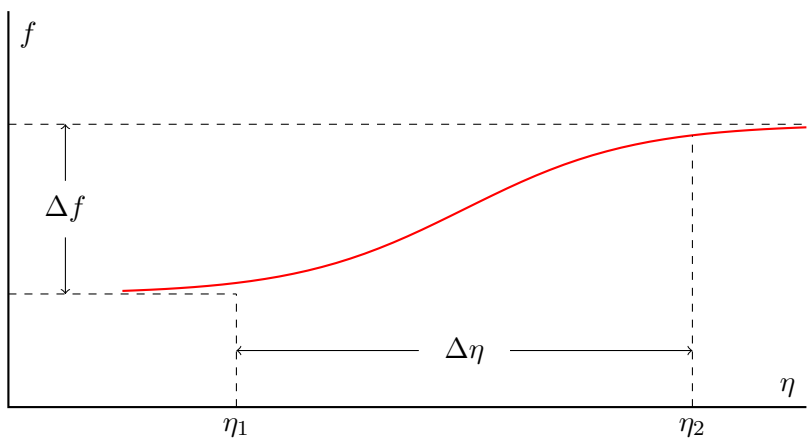
\begin{figure}[htp]
	\centering	
	\begin{tikzpicture}[scale=1.5]
	\draw [thick] (7,0) -- (0,0) -- (0,3.5);
	\draw[thick, red, smooth, domain=1:7] plot (\x, {.75*tanh((\x - 4)/1.5) + 1.75});
	\draw [dashed] (0,1) -- (2,1);
	\draw [dashed] (0,2.5) -- (7,2.5);
	\draw [dashed] (2,0) -- (2,1);
	\draw [dashed] (6,0) -- (6,2.4);
	\node [below right] at (0,3.5) {\small $f$};
	\node [above left] at (7,0) {\small $\eta$};
	\draw [->] (.5,2) -- (.5,2.5);
	\node at (.5,1.75) {\small $\Delta f$};
	\draw [->] (.5,1.5) -- (.5,1);
	\draw [->] (3.6,.5) -- (2,.5);
	\node at (4,.5) {\small $\Delta \eta$};
	\draw [->] (4.4,.5) -- (6,.5);
	\node [below] at (2,0) {\small $\eta_1$};
	\node [below] at (6,0) {\small $\eta_2$};
	\end{tikzpicture}
	\caption{Evolution of the form (\ref{f}). \label{fig:f}}
\end{figure}

\section{Amplification inside the Hubble radius}
\label{sec:prior}

In this paper, we are interested in general upper limits imposed by the back-reaction  and Schwinger effect on the possible strength of magnetic field on a particular comoving spatial scale characterized by wavenumber $k$.  We can start amplifying the field prior to Hubble-radius crossing during inflation. A model for such amplification was proposed in our works \cite{Shtanov:2019civ, Shtanov:2019gpx}.  Assuming constant (slowly evolving) kinetic coupling $I$, suppose that the chiral coupling evolves as (see figure \ref{fig:f}) 
\begin{equation}\label{f}
	f (\eta) = \frac12 \Delta f \tanh \left( \frac{2 \eta - \eta_1 - \eta_2}{\Delta \eta} \right) + \text{const} \, ,
\end{equation}
in which $\Delta \eta = \eta_2 - \eta_1 > 0$ is its temporal width.  Then $f'_0 \equiv f'\left((\eta_1+\eta_2)/2\right) = \Delta f / \Delta \eta$. 

Introducing the dimensionless time $\tau = (2 \eta - \eta_1 - \eta_2) / \Delta \eta$ and wavenumber $p = k \Delta \eta / 2$, one can write the general solution of (\ref{eqAh}) (with zero electric conductivity) in terms of the Ferrers functions \cite[Chapter~14]{Olver}:
\begin{equation} \label{solA}
	\cA_h ( \tau ) = c_+ \rP_\nu^\mu \left( \tanh \tau \right) + c_- \rP_\nu^{-\mu} \left( \tanh \tau \right) \, ,	
\end{equation}
with
\begin{equation} \label{munu} 
	\mu = \ri p \, , \qquad \nu = q - \frac12  \, , \qquad q \equiv \frac12 \sqrt{1 + 2 h \Delta f p } \, . 
\end{equation}

The asymptotics of the canonically normalized quantity $I \cA_h \sim e^{- \ri k \eta} = e^{- \ri p \tau}$ as $ \tau \to - \infty$ determines the constants $c_+$ and $c_-$ in (\ref{solA}).  By considering the opposite asymptotics $I \cA_h \sim \alpha_k e^{- \ri p \tau} + \beta_k e^{\ri p \tau} $ as $ \tau \to \infty$ in (\ref{solA}), one reads off the Bogolyubov's coefficients $\alpha_k$ and $\beta_k$.  Skipping a simple calculation, we present here the result: 
\begin{equation}
	\alpha_k = \frac{\Gamma (1 - \ri p) \Gamma (- \ri p)}{\Gamma \left( \frac12 + q - \ri p \right) \Gamma \left( \frac12 - q - \ri p \right)} \, , \qquad
	\beta_k = - \ri \frac{\cos ( \pi q )}{\sinh (\pi p ) } \, ,
\end{equation}
with the required property $\left| \alpha_k \right|^2 - \left| \beta_k \right|^2 = 1$.  
After the evolution of the coupling, the mean number of quanta in a given mode is 
\begin{equation}\label{nk}
	n_k = \left| \beta_k \right|^2 = \frac{\cos^2 (\pi q )}{\sinh^2 (\pi p )} \, .
\end{equation}

\begin{figure}[htp]
	\centering
	\includegraphics[width=.7\textwidth]{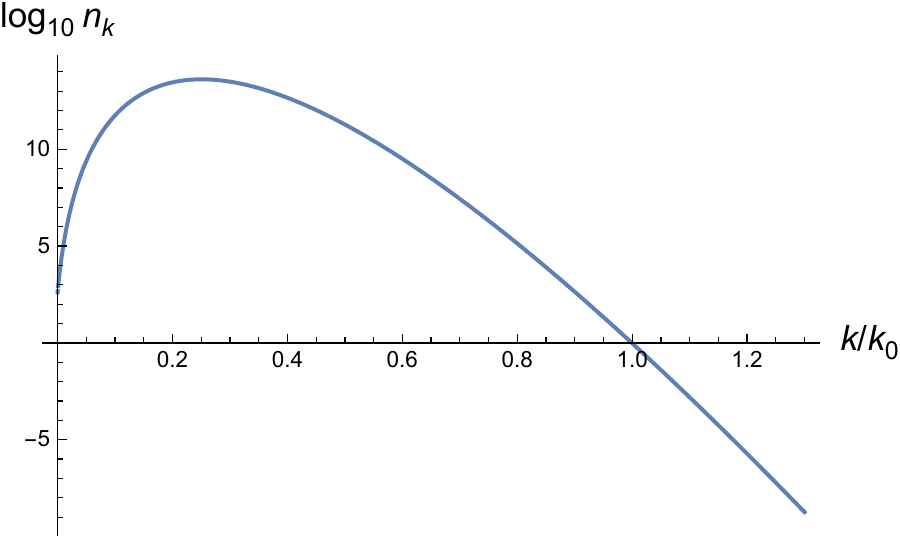}
	\caption{Spectrum (\ref{nk}) on a logarithmic scale for the helicity satisfying $h \Delta f < 0$ and for $|\Delta f| = 40$. \label{fig:plot}}
\end{figure}

For the helicity satisfying $h \Delta f < 0$, the quantity $q$, given by (\ref{munu}), is purely imaginary for $p > 1 / 2 |\Delta f|$.  In the approximation $p \gg \text{max}\,\left\{ {1}/{2 |\Delta f|}\, , \, {1}/{\pi} \right\}$, we then obtain
\begin{equation} \label{n}
	n_k \approx e^{2 \pi \left( \sqrt{|\Delta f| p / 2} - p \right)} = e^{\pi | \Delta f | \left( \sqrt{ k / k_0} - k / k_0 \right) } \, ,
\end{equation}
where $k_0 = \left| f_0' \right|$ is the wavenumber at which spectrum (\ref{n}) reaches unity on the slope of its exponential decline.  The exponent of this expression reaches maximum at $k_{\rm m} = k_0 / 4$, with the maximum mean occupation number $n_{\rm m} = e^{\pi | \Delta f| / 4}$, which is exponentially large for $| \Delta f | \gg 1$. Spectrum (\ref{nk}) is plotted in figure \ref{fig:plot} on a logarithmic scale for a typical value $|\Delta f| = 40$ (see section \ref{sec:eff}).  The mean occupation numbers for the opposite helicity can be neglected. 

The spectral densities (\ref{specB}) and (\ref{specE}) are of comparable magnitudes, and, since one of the helicities is dominating in $n_k$, we have, using (\ref{n}), 
\begin{equation}\label{PBE}
	\cP_B \approx \cP_E \approx \frac{k^4}{4 \pi^2 a^4 I^2} \left[ 1 + e^{\pi | \Delta f | \left( \sqrt{ k / k_0} - k / k_0 \right) } \right] \, .
\end{equation}

We observe that the spectral densities are peaked at the central value $k = k_{\rm m}$ with width $\Delta k \simeq k_{\rm m} / \sqrt{|\Delta f|} \ll k_{\rm m}$ for $|\Delta f| \gg 1$.  Thus, electric and magnetic fields are generated in this scenario with similar spectra in the spectral region of amplification.

Using a Gaussian approximation to (\ref{PBE}), one can estimate the total excess (over vacuum) of the electromagnetic energy density and the electromagnetic field strength after the end of the process of amplification:
\begin{equation} \label{main}
\rho_{\rm em} \simeq \frac{2 k_{\rm m}^4}{\pi^2 a^{4}} \frac{e^{\pi | \Delta f |/4} }{| \Delta f|^{1/2}} \, , \qquad B \simeq E \simeq \frac{\sqrt{\rho_{\rm em}}}{I} \, .
\end{equation}

The theory contains two free parameters, $\Delta f$ and $\Delta \eta$, which can be adjusted to produce maximally helical electromagnetic fields of desirable strength with spectral density centred at a desirable wavenumber $k_{\rm m} = | \Delta f | / 4 \Delta \eta$.

The mechanism of electromagnetic-field generation considered here is quite general and can take place at any stage of cosmological evolution where electric conductivity can be neglected. One can use it to amplify electromagnetic field at the inflationary stage prior to the Hubble-radius crossing for the relevant comoving spatial scale. 

Solution (\ref{solA}) for the electromagnetic field at the centre of the spectral domain of amplification and in the interval $\eta_1 < \eta < \eta_2$ grows exponentially: $\cA_h \propto e^{k_\rmm \eta}$. Assuming $|k_\rmm \eta_2| \lesssim 1$ and taking into account the prefactors $\propto a^{-4}$ in (\ref{specB}) and (\ref{specE}), we conclude that the peak of the electromagnetic spectral density is reached at the time where $|k_\rmm \eta| = k_\rmm / a H \approx 2$, i.e., a little before the Hubble-radius crossing.\footnote{And not at the end of inflation, as was mistakenly assumed in \cite{Shtanov:2019civ, Shtanov:2019gpx} with ensuing gross overestimation of the resulting magnetic field.} 

\section{Back-reaction and Schwinger effect}
\label{sec:eff}

Regardless of a concrete model of electromagnetic-field amplification, its characteristics should meet the constraints imposed by the considerations of back-reaction and Schwinger effect. 

The back-reaction constraint means that the electromagnetic energy density $\rho_{\rm em} = I^2 \left( B^2 + E^2 \right)/ 2$ at least should not exceed the energy density of the rest of matter $\rho = M_\rP^2 H^2$, where $M_\rP = \sqrt{3 / 8 \pi G} \approx 4.2 \times 10^{18}\, \text{GeV}$ is the reduced Planck mass expressed through the Newton's gravitational constant $G$. Hence, we should have
\begin{equation}\label{back}
	I (E + B) \lesssim M_\rP H \, . 
\end{equation}
The electromagnetic energy density at the time of Hubble-radius crossing is estimated as $\rho_{\rm em} \simeq Q^2 H^4$ with the requirement $Q \lesssim M_\rP / H$ for the amplification factor of electromagnetic field relative to the purely vacuum fluctuations on the Hubble scale. In the scenario of amplification considered in the previous section, from (\ref{main}) taken at the Hubble-radius crossing, we have
\begin{equation}\label{Q}
	Q \simeq \frac{e^{\pi | \Delta f |/8}}{| \Delta f|^{1/4}} \lesssim \frac{M_\rP}{H} \, .
\end{equation}
For a typical value $H \simeq 10^{-6} M_\rP$, in the scenario of the preceding section, this requirement is translated as $|\Delta f| \lesssim 40$, which limiting value was chosen in the plot of figure \ref{fig:plot}. 

Consider now the role of electric conductivity. For our model of amplification from the previous section, the second term in eq.~(\ref{eqAh}) can be neglected under the condition
\begin{equation}\label{schw}
	\frac{a \sigma}{I^2} \ll \left| \frac{\cA_h''}{\cA_h'} \right| \lesssim k_\rmm \quad \Rightarrow \quad \frac{\sigma}{I^2} \ll H \, ,
\end{equation}
since $k_\rmm = 2 a H$ at the peak of electromagnetic field strength. In the opposite case, $\sigma/I^2 \gg H$, the term with conductivity in eq.~(\ref{eqAh}) will cause the derivative $\cA_h'$ to exponentially decrease on time scale much exceeding the Hubble time, making the mode $\cA_h$ approximately constant in time.

The main cause of electric conductivity during inflation is the Schwinger effect of production of charged particle-antiparticle pairs from the vacuum. For fermionic particles with mass $m$ and electric charge $q$, the conductivity calculated in the approximation of homogeneous and slowly changing electromagnetic fields is given by \cite{Kobayashi:2014zza, Hayashinaka:2016qqn, Bavarsad:2016cxh, Domcke:2019qmm}
\begin{equation}\label{conduc}
	\sigma \simeq \frac{q^3}{6 \pi^2} \frac{B}{H} \coth \left( \frac{\pi B}{E} \right) e^{- \pi m^2 / q E} \geq \frac{q^3}{6 \pi^3} \frac{E}{H} e^{- \pi m^2 / q E} \, ,
\end{equation}
where we have used the inequality $x \coth x \geq 1$.

During inflation, the fermion mass $m$ is determined by its Yukawa coupling and by the fluctuations of the Higgs field $\phi \sim H$, so that $m = \gamma H$, where $\gamma \sim 10^{-6}$ for the electron.  Hence, inequality (\ref{schw}) together with (\ref{conduc}) leads to the constraint
\begin{equation}
	\frac{q^3}{6 \pi^3} \frac{E}{H^2} e^{- \pi \gamma^2 H^2 / q E} \lesssim I^2 \, .
\end{equation}
For the uppermost admissible value of $E/H^2 \simeq 6 \pi^3 I^2/q^3$, the exponent in this expression can be neglected, and we obtain the bound 
\begin{equation}\label{schwc}
	E \lesssim \frac{6 \pi^3}{q^3} I^2 H^2 \simeq 10^3 I^2 H^2 \, .
\end{equation}
where we have used $q \approx \sqrt{4 \pi / 137} \approx 0.3$ for the unit charge.

In view of (\ref{back}) and (\ref{schwc}), in the scenario of the preceding section, we then have
\begin{equation}\label{back-abs}
	B \simeq E \lesssim \text{min}\, \left \{ \frac{M_\rP H}{I} \, , \, 10^3 I^2 H^2 \right \} \leq 10 M_\rP^2 \left( \frac{H}{M_\rP} \right)^{4/3} 	\, .
\end{equation}

Constraint (\ref{back-abs}) at the Hubble-radius crossing would be insufficient to obtain reasonable magnetic fields on large spatial scales without further amplification. Indeed, if we assume adiabatic evolution $B \propto a^{-2}$ until the present time, then, for the current magnetic field, we obtain an upper bound
\begin{align}\label{BH}
	B_0 &\lesssim 10 M_\rP^2 \left( \frac{H_k}{M_\rP} \right)^{4/3} \left( \frac{a_k}{a_0} \right)^2 = 10 \left( \frac{M_\rP}{H_k} \right)^{2/3} \left( \frac{k}{a_0} \right)^2 \nonumber \\ &\sim \frac{400}{\lambda^2} \left( \frac{M_\rP}{H_k} \right)^{2/3} \sim 10^{-54} \left( \frac{M_\rP}{H_k} \right)^{2/3} \left( \frac{\text{Mpc}}{\lambda} \right)^2 \, \text{G} \, .
\end{align} 
Here, $a_k = k / H_k$ and $H_k$ are the scale factor and Hubble parameter, respectively, at the Hubble-radius crossing during inflation, and $\lambda = 2 \pi a_0 / k$ is the present value of the comoving spatial scale corresponding to the wavenumber $k$. 

Disengaging from the specific scenario of the preceding section, we can contemplate generation of the maximal possible magnetic field at the Hubble-radius crossing compatible with the requirements of weak back-reaction ($I B \lesssim M_\rP H$) and weak coupling ($I \gtrsim 1$). With no subsequent amplification, instead of (\ref{BH}) we then obtain the estimate
\begin{equation}\label{BH1}
	B_0 \lesssim M_\rP H_k \left( \frac{a_k}{a_0} \right)^2 \sim 10^{-55}\, \frac{M_\rP}{H_k} \left( \frac{\text{Mpc}}{\lambda} \right)^2 \, \text{G} \, . 
\end{equation}

Such low values (\ref{BH}) and (\ref{BH1}) of the resulting magnetic field are insufficient for successful magnetogenesis in scenarios with reasonable energy scale of inflation.  Hence, further amplification of magnetic field would be required on given comoving spatial scales after Hubble-radius crossing.

\section{Amplification outside the Hubble radius}
\label{sec:after}

Amplification of electromagnetic field outside the Hubble radius can be done by evolving the coupling $I$ in (\ref{Lem}), and we assume the usual conditions
\begin{equation}
\left\{ k^2 , \, k \frac{\left| \left( I^2 f \right)'\right| }{I^2} \right\} \ll 	\frac{\left| I'' \right|}{I} 
\end{equation}
to be valid well after Hubble-radius crossing. Under these conditions, the terms containing $k$ in (\ref{eqAh1}), hence, also in (\ref{eqAh}) can be neglected. Dropping also the terms with electric conductivity, we observe that equation (\ref{eqAh}) reduces to $\left( I^2 \cA'_h \right)' = 0$, and its general solution is given by
\begin{equation}\label{Asol}
	\cA_h \simeq \frac{1}{I_k} \left[ Q_1 + Q_2 k \int \left( \frac{I_k}{I} \right)^2 d \eta \right] = \frac{1}{I_k} \left[ Q_1 + Q_2 \int \left( \frac{I_k}{I} \right)^2 R_k\, d N \right] \, .
\end{equation}
Here, integration is taken from the moment of Hubble-radius crossing, $Q_1$ and $Q_2$ are integration constants that reflect evolution of the mode prior to Hubble-radius crossing, and, in the last expression, the integration variable was transformed as $k d \eta = k d a / a^2 H = k d N / aH = R_k\, d N$, where $R_k = k / a H < 1$ and $N = \ln a$ is the number of $e$-foldings.\footnote{Note that, even if the coupling $I$ stops evolving, mode (\ref{Asol}) can still continue to grow as $\cA_h \propto R_k \propto 1/a H$, with magnetic field evolving as $B \propto |\cA_h|/a^2 \propto 1 / a^3 H$, which law was noted and used in \cite{Kobayashi:2014sga, Kobayashi:2019uqs, Fujita:2019pmi}.}
  
Focusing on the modes with a particular wavenumber $k$, we would like to obtain the largest possible amplitude for $\cA_h$ in the end.  The process of field amplification is limited by the considerations of back-reaction and Schwinger effect, similar to those discussed in the previous section. The back-reaction constraints have the form (\ref{back}). As regards the Schwinger effect, we estimate the related conductivity $\sigma$ at all stages by expression in (\ref{conduc}). This can be justified, e.g., by integrating \cite[eq.~(4.11)]{Domcke:2019qmm}, although the real situation with the Schwinger current, especially after inflation, is not very clear and may be more complicated \cite{Gorbar:2019fpj, Sobol:2020frh} (see also discussion in section \ref{sec:sum}). Then, upper bound of the form (\ref{schwc}) is applied for the electric field at all stages of cosmological evolution.  We thus have the following constraints: 
\begin{align}
	I E \lesssim M_\rP H \quad &\Rightarrow \quad \abs{Q_2} \frac{k^2}{a^2} \frac{I_k}{I} \lesssim M_\rP H \, , \label{Ebr} \\
	I B \lesssim M_\rP H \quad &\Rightarrow \quad \frac{k^2}{a^2} \frac{I}{I_k} \left[ \abs{Q_1} + \abs{Q_2} \int \left( \frac{I_k}{I} \right)^2 R_k\, d N \right]  \lesssim M_\rP H \, , \label{Bbr} \\
	E \lesssim 10^3 I^2 H^2 \quad &\Rightarrow \quad \abs{Q_2} \frac{k^2}{a^2} \frac{I_k}{I} \lesssim 10^3 I^3 H^2 \, . \label{Schw}
\end{align}
These constraints take into account only contribution to the fields from the modes around the specified wave number $k$; they are quite conservative in this sense.  The role of the whole spectrum will be discussed in section \ref{sec:spectrum}.

Conditions (\ref{Ebr}) and (\ref{Schw}) are equivalent to the following respective constraints for the integrand in the last expression of (\ref{Asol}):
\begin{align}
	\left( \frac{I_k}{I} \right)^2 R_k &\lesssim \frac{M_\rP^2}{\abs{Q_2}^2 H^2} R_k^{-3} \, , \label{Ebr1} \\
	\left( \frac{I_k}{I} \right)^2 R_k &\lesssim \frac{\left( 10 I_k \right)^{3/2}}{\abs{Q_2}^{1/2}} \label{Schw1} \, .
\end{align}
It is notable that the right-hand side of (\ref{Ebr1}) monotonically increases with time, while the right-hand side of (\ref{Schw1}) is constant. If the constant threshold on the right-hand side of (\ref{Schw1}) is reached by the integrand before the conductivity from preheating becomes significant, then it is the Schwinger effect that stops the growth of the mode $\cA_h$ by creating plasma with sufficiently high conductivity.\footnote{We note that the Schwinger particle production in this type of models of cosmological magnetogenesis has even been contemplated as an alternative to traditional preheating \cite{Tangarife:2017rgl, Gorbar:2019fpj}.} In the opposite case, it is electric conductivity from preheating that stops the growth of the mode. 

We are not interested in scenarios without significant amplification of magnetic field outside the Hubble radius because, under the condition of weak gauge coupling, they lead to a moderate upper bound (\ref{BH1}). Hence, for our purposes, we can assume that the main contribution to the final amplitude $\cA_h$ comes from the integral part in (\ref{Asol}), and constraints on the constant $Q_1$ will be of no interest to us. The integrand in (\ref{Asol}) reaches its maximal value on the time interval before electric conductivity from reheating becomes significant, and this gives an estimate for the final amplitude up to a factor of order unity:
\begin{equation}\label{estim}
	\cA_h \simeq \frac{Q_2}{I_k} \int \left( \frac{I_k}{I} \right)^2 R_k\, d N \sim \frac{Q_2}{I_k} \left[ \left( \frac{I_k}{I} \right)^2 R_k \right]_\max = Q_2 I_k \left( \frac{R_k}{I^2} \right)_{\max} \, ,
\end{equation}
where the label `max' indicates the moment of time where the maximum of the integrand is reached on the specified time interval. The integrand is bounded from above by non-decreasing functions of time (\ref{Ebr1}) and (\ref{Schw1}), whence we have
\begin{equation}\label{int}
\abs{\cA_h}	\lesssim \min \left\{ \frac{1}{\abs{Q_2} I_k} \left( \frac{M_\rP^2}{H^2} R_k^{-3} \right)_\rr\, , \ 10^{3/2} \left( \abs{Q_2} I_k \right)^{1/2} \right\} \, ,
\end{equation}  
where the label `r' denotes quantities evaluated at the moment of reheating.   

Looking at (\ref{int}), we observe that the absolute maximal upper bound for $\abs{\cA_h}$ is obtained for 
\begin{equation}\label{ineq}
	\abs{Q_2} I_k \simeq 10^{-1} \left[ \left( \frac{M_\rP}{H} \right)^{4/3} R_k^{-2} \right]_\rr \, ,
\end{equation}
and is given by
\begin{equation}\label{Amax}
	\abs{\cA_h} \lesssim 10 \left[ \left( \frac{M_\rP}{H} \right)^{2/3} R_k^{-1} \right]_\rr \, .  
\end{equation}

The remaining constraint (\ref{Bbr}) can be shown to imply $\left. R_k \right|_\rr = k / a_\rr H_\rr \lesssim 1$, i.e., our estimates are valid for the modes that are outside the Hubble radius by the moment of reheating. Also note that the value of $I$ is bounded by (\ref{Ebr1}) and (\ref{ineq}), so that
\begin{equation}
	I \gtrsim \abs{Q_2} I_k \frac{H}{M_\rP} R_k^2 \gtrsim \abs{Q_2} I_k \left( \frac{H}{M_\rP} R_k^2 \right)_\rr \simeq 10^{-1} \left( \frac{M_\rP}{H_\rr} \right)^{1/3} \gtrsim 1
\end{equation}
provided $H_\rr \lesssim 10^{-3} M_\rP$. Thus, we are in a safe weak-coupling regime for our gauge field.

The current strength of magnetic field on the scale of interest is estimated from (\ref{specB}) and (\ref{Amax}) as\footnote{Assuming the law $B \propto a^{-2}$ in the hot universe, we neglect the possible chiral and turbulence effects \cite{Banerjee:2004df, Boyarsky:2011uy, Kahniashvili:2012uj, Saveliev:2013uva, Boyarsky:2015faa, Hirono:2015rla, Subramanian:2015lua, Gorbar:2016qfh, Sidorenko:2016vom, Gorbar:2016klv, Pavlovic:2016gac, Brandenburg:2017rcb, Rogachevskii:2017uyc, Schober:2017cdw} that may modify the evolution of the power spectrum.}
\begin{equation}
	B_0 \simeq \frac{k^2}{2 \sqrt{2} \pi a_0^2} \left| \cA_h \right| \lesssim \left( \frac{k}{a_0} \right) \left( \frac{a_\rr}{a_0} \right) M_\rP^{2/3} H_\rr^{1/3} \simeq \frac{k T_0}{a_0} g_\rr^{-1/6} \left( \frac{M_\rP}{T_\rr} \right)^{1/3} \, ,
\end{equation}
where $g_\rr$ is the number of relativistic degrees of freedom in thermal equilibrium after reheating ($g_\rr \approx 100$ in the Standard Model), $T_\rr$ is the reheating temperature, and $T_0 = 2.34 \times 10^{-4}\, \text{eV}$ is the current temperature of relic photons.  Substituting here the physical numbers, we get
\begin{equation}\label{mainf}
	B_0 \lesssim 10^{-30} g_\rr^{-1/6} \left( \frac{M_\rP}{T_\rr} \right)^{1/3} \frac{\text{Mpc}}{\lambda}\, \text{G} \, .
\end{equation}

It is instructive to see whether the Schwinger effect played a significant role in our upper bound. The final amplitude for the mode for fixed $Q_2$ and $I_k$ is estimated by (\ref{estim}). Now, condition (\ref{Ebr1}) implies the inequality $\abs{Q_2} I_k \lesssim \left. M_\rP I / H R_k^2 \right|_\max$, using which, we get 
\begin{equation}\label{Amaxns}
	\abs{\cA_h} \lesssim \left( \frac{M_\rP}{I H} R_k^{-1} \right)_\max \lesssim \left( \frac{M_\rP}{H} R_k^{-1} \right)_\rr \, ,
\end{equation}
where we have taken into account the weak-coupling constraint $I \gtrsim 1$.  Condition (\ref{Bbr}) gives a weaker constraint compared to (\ref{Amaxns}). 

The result (\ref{Amaxns}) is larger than (\ref{Amax}) by a factor $10^{-1} \left( M_\rP / H_\rr \right)^{1/3}$, which is significant for low temperatures of reheating.  The estimate for the current magnetic field in this case is
\begin{equation}\label{ns}
	B_0 \lesssim 10^{-31} g_\rr^{-1/3} \frac{M_\rP}{T_\rr} \frac{\text{Mpc}}{\lambda}\, \text{G} \, .
\end{equation}
This estimate would be applicable if the Schwinger particle production were suppressed, e.g.,  by enhancement of the masses of charged particles (perhaps, by some non-trivial coupling to $I$), as discussed in \cite{Kobayashi:2019uqs}.

\section{The role of spectrum}
\label{sec:spectrum}

In the previous section, we derived the weakest possible upper bounds, considering electromagnetic field only around one particular scale $k$. If we take into account that magnetogenesis, in fact, takes place on all spatial scales, then our estimates may be strengthened.  Let us show that this happens only mildly, i.e., that estimate (\ref{mainf}) can be saturated by order of magnitude.  We demonstrate this by constructing a scenario in which the spectrum for electric field remains to be peaked around one particular scale $k_\rmm$.

The fastest possible growth of the integrand in (\ref{Asol}) compatible with the back-reaction constraint (\ref{Ebr}) during inflation occurs under the law\footnote{It is remarkable that evolution of the form (\ref{Iev}) is also obtained in numerical simulations in models of type (\ref{Lem}) in which back-reaction of electric field on the inflaton dynamics is taken into account through the dependence of the coupling $I$ on the inflaton field $\phi$ \cite{Sobol:2018djj}.  In models of this kind, back-reaction becomes important as the energy density of electric field reaches the value of the order $\rho_E \sim \epsilon \rho_{\rm inf}$, where $\epsilon = M_\rP^2 \left[ V' (\phi) / V (\phi) \right]^2$ is the small inflationary slow-roll parameter, and $V (\phi)$ is the inflaton potential. Back-reaction effects of this kind, which are model-dependent and which we do not consider in this paper, would further reduce our upper bounds for the admissible electric (hence, also magnetic) field.} 
\begin{equation}\label{Iev}
	I \propto a^{-2} \propto \eta^2 \, .
\end{equation}
The solution for the mode is then
\begin{equation}\label{aampl}
	\cA_h \propto \int \frac{d \eta}{I^2} = C_1 + C_2  a^3 \, .
\end{equation}
The fields in the growing mode evolve as $E \propto a^2$ and $B \propto E/a \propto a$, growing with time.

Consider then scenario (\ref{Iev}), (\ref{aampl}) with preliminary amplification as described in section \ref{sec:prior}. This evolution of $\cA_h$ takes place for all super-Hubble modes starting from the moment $\eta_\rmm$ of the beginning of evolution of $I$, which we take to be the time of Hubble-radius crossing of the mode with wavenumber $k_\rmm$. By matching the initial conditions at $a_\rmm$, we will have
\begin{align}\label{long}
	\cA_h &\simeq \cA_\rmm \left[ 1 + \frac{k}{3 k_\rmm} \left( \frac{a}{a_\rmm} \right)^3 \right] = \cA_\rmm \left[ 1 + \frac{k}{3 k_\rmm} \left( \frac{a H}{k_\rmm} \right)^3 \right] \, , \\ 
	\frac{\cA_h'}{k} &\simeq \cA_\rmm \left( \frac{a H}{k_\rmm} \right)^4 \, , \qquad k \lesssim k_\rmm \, ,
\end{align}
where $\cA_\rmm = \cA_m \left (\eta_\rmm, k \right)$ is the mode at the moment $\eta_\rmm$, and $k_\rmm = H a_\rmm$.  

As regards the modes with $k \gtrsim k_\rmm$, the amplitude $|I \cA_h|$ remains constant till the Hubble-radius crossing, where $a_k = k / H$, so that $\cA_h \propto 1/I \propto a^2$, after which it evolves as (\ref{aampl}). Hence,
\begin{align}
	\cA_h &\simeq \cA_\rmm \left( \frac{a_k}{a_\rmm} \right)^2 \left( \frac{a}{a_k} \right)^3 = \cA_\rmm \frac{(a H)^3}{k_\rmm^2 k} \, , \\ \frac{\cA_h'}{k} &\simeq 3 \cA_\rmm \frac{(a H)^4}{k_\rmm^2 k^2} \, , \qquad k \gtrsim k_\rmm \, , \label{short}
\end{align}

Equations (\ref{long})--(\ref{short}) can be interpolated for all $k$ as
\begin{align}
	\cA_h &\simeq \cA_\rmm \left[1 + \frac{k/k_\rmm}{3 + (k/k_\rmm)^2} \left( \frac{a H}{k_\rmm} \right)^3 \right] \, , \\ \frac{\cA_h'}{k} &\simeq \cA_\rmm \frac{1}{1 + k^2 / 3 k_\rmm^2} \left( \frac{a H}{k_\rmm} \right)^4 \, .
\end{align}
The spectral densities during inflation are then estimated as
\begin{align}
	\cP_B &= \frac{d \left\langle B^2 \right\rangle}{d \ln k} \simeq \frac{k^4}{8 \pi^2 a^4} \left[1 + \frac{k/k_\rmm}{3 + (k/k_\rmm)^2} \left( \frac{a H}{k_\rmm} \right)^3 \right]^2 \sum_h \left| \cA_h (\eta_\rmm, k) \right|^2 \, , \label{sspecB} \\
	\cP_E &= \frac{d \left\langle E^2 \right\rangle}{d \ln k} \simeq \frac{k^4}{8 \pi^2 a^4} \frac{1}{\left(1 + k^2 / 3 k_\rmm^2\right)^2} \left( \frac{a H}{k_\rmm} \right)^8 \sum_h \left| \cA_h (\eta_\rmm, k) \right|^2 \, . \label{sspecE}
\end{align}
Due to the strong narrow peak of $Q (k) \equiv \left| I_k \cA_h (\eta_k, k) \right| \approx \left( 1 + n_k \right)^{1/2}$ (see eqs.~(\ref{n}) and (\ref{PBE})) at the Hubble-radius crossing, the spectrum for electric field remains to be peaked at $k = k_\rmm$ and is flat in the region $k \gg k_\rmm$.  Magnetic field is suppressed on all relevant scales $k \lesssim a H$ with respect to the electric one. This, then, does not strengthen much our estimates made in the previous section. 

We also remark that, although our upper bounds (\ref{Amax}) and (\ref{Amaxns}) are independent of the value of $Q (k)$ at that particular wavenumber, the role of initial amplification inside the Hubble radius is significant because it shapes the initial power spectrum, allowing it to be peaked at a given wavenumber and thus making our upper bounds virtually the least upper bounds.

\section{Summary and discussion}
\label{sec:sum}

In this paper, we have derived an upper bound (see (\ref{mainf})) 
\begin{equation}\label{maind}
	B_0 \lesssim 10^{-30} g_\rr^{-1/6} \left( \frac{M_\rP}{T_\rr} \right)^{1/3} \frac{\text{Mpc}}{\lambda}\, \text{G} 
\end{equation}
for the outcome of inflationary and post-inflationary magnetogenesis based on Lagrangian (\ref{Lem}) under weak coupling $I \gtrsim 1$.  In doing this, we took into account the constraints from back-reaction and Schwinger effect. Our upper bound is conservative in that it takes into account only electromagnetic fields on a particular spatial scale of interest and can only be lowered by considering the whole power spectrum. Still, we have shown that there exists a scenario in which the power spectrum is peaked around the spatial scale of interest, which makes our bound, in some sense, virtually the least upper bound. Our model-independent result (\ref{maind}) is consistent with the estimates obtained in the literature for specific models (e.g., in \cite{Kobayashi:2014zza, Kobayashi:2019uqs}).

The upper bound (\ref{maind}) allows one to appreciate the difficulty in producing sufficiently large magnetic fields on large spatial scales. Indeed, the lowest possible temperature of reheating lies in the MeV range \cite{Hannestad:2004px}, and even for such low reheating temperature, one gets $B_0 \lesssim 10^{-23}\,\text{G}$ on the megaparsec scale. Estimate (\ref{maind}) is then valid for comoving spatial scales larger than the Hubble radius at this temperature, i.e., for $\lambda \gtrsim 100\,\text{pc}$.  

In our treatment, we used the simple expression (\ref{conduc}) for the electric conductivity induced by the Schwinger effect at the post-inflationary stage. Hoping that it gives correct estimates by order of magnitude, we should point out that recent elaborate calculations in the kinetic approach reveal an oscillatory and non-Markovian character of the electric current and field after inflation, failing to satisfy the simple Ohm's law and decaying slower than in the case of na\"{\i}ve application of (\ref{conduc}) \cite{Gorbar:2019fpj, Sobol:2020frh}.  This may call for revisions of our estimates in more refined treatments of the Schwinger effect.

If the Schwinger effect for some reason is suppressed in the early universe (e.g., because the masses of charged particles are sufficiently enlarged by some non-trivial mechanism), then only back-reaction is important, and our estimate for the final magnetic field becomes (see (\ref{ns}))
\begin{equation}\label{mainw}
	B_0 \lesssim 10^{-31} g_\rr^{-1/3} \frac{M_\rP}{T_\rr} \frac{\text{Mpc}}{\lambda}\, \text{G} \, .
\end{equation}
This would allow one to obtain $B_0 \sim 10^{-15}\,\text{G}$ on the megaparsec scale by choosing $T_\rr \sim 10^2\,\text{GeV}$. In some specific scenarios, generation of magnetic fields of this order of magnitude requires even lower reheating temperatures, of the order of GeV \cite{Fujita:2016qab, Fujita:2019pmi}.

We did not take into account the contribution of electromagnetic field to the primordial power spectrum of energy-density fluctuations, but the corresponding constraints on the magnetic field \cite{Markkanen:2017kmy} appear to be much weaker than those obtained in this paper.

Finally, we note that primordial helical hypermagnetic fields may also be responsible for generating baryon asymmetry of the universe \cite{Bamba:2006km, Anber:2015yca, Fujita:2016igl, Kamada:2016eeb, Kamada:2016cnb, Jimenez:2017cdr, Domcke:2019mnd}.  This imposes a typical constraint \cite{Kamada:2016cnb, Domcke:2019mnd}
\begin{equation}\label{bacon}
	B_0 \lesssim 10^{-21}\, \left( \frac{\text{Mpc}}{\lambda} \right)^{1/2} \, \text{G} 
\end{equation}
on the admissible values of $B_0$ on the spatial scale $\lambda$ for the fields of {\em maximal\/} helicity that existed prior to the electroweak transition.  This constraint is weaker than our upper bound (\ref{maind}) for not very small spatial scales. 

\section*{Acknowledgments}

We are grateful to Eduard Gorbar for valuable comments and discussion. Y.~S.\@ acknowledges support from the National Academy of Sciences of Ukraine in frames of priority project ``Fundamental properties of matter in the relativistic collisions of nuclei and in the early Universe'' (No.~0120U100935) and from the scientific program ``Astronomy and Space Physics'' (project 19BF023-01) of the Taras Shevchenko National University of Kiev.

\end{document}